\documentclass[a4paper,fleqn]{cas-dc}
\usepackage[numbers]{natbib}
\usepackage{multirow}
\usepackage[utf8]{inputenc}
\usepackage{graphicx}
\usepackage{hyperref}
\usepackage{epstopdf}
\usepackage{adjustbox}
\usepackage{float}
\usepackage{cancel}

\def\tsc#1{\csdef{#1}{\textsc{\lowercase{#1}}\xspace}}
\tsc{WGM}
\tsc{QE}
\tsc{EP}
\tsc{PMS}
\tsc{BEC}
\tsc{DE}
\DeclareUnicodeCharacter{2212}{-}
\begin{document}
\let\WriteBookmarks\relax
\def\floatpagepagefraction{1}
\def\textpagefraction{.001}
\shorttitle{Thermally assisted magnetization reversal of a magnetic nanoparticle driven by a down-chirp microwave field pulse}
\shortauthors{M. T. Islam et~al.}

\title [mode = title]{Thermally assisted magnetization reversal of a magnetic nanoparticle driven by a down-chirp microwave field pulse }

\author[1]{M. T. Islam}[orcid=0000-0002-3846-4009]

\author[1]{M. A. J. Pikul}[orcid=0000-0002-7500-1572]

\author[2]{X. S. Wang}[orcid=0000-0003-1374-4407]
\cormark[1]
\ead{justicewxs@hnu.edu.cn}


\address[1]{Physics Discipline, Khulna University, Khulna 9208, Bangladesh}

\address[2]{School of Physics and Electronics, Hunan University, Changsha 410082, China}


		
\cortext[cor1]{Corresponding author: }
\begin{abstract}
   It has been shown that a single-domain magnetic nanoparticle can be effectively switched by a linear down-chirp microwave field pulse (DCMWP) in zero temperature limit. However, finite temperature is ubiquitous in practice. Here, we study the effect of finite temperature on the DCMWP-induced magnetization reversal based on the stochastic Landau-Lifshitz-Gilbert equation. It is found that any one of the three controlling parameters of a DCMWP, i.e. the amplitude, chirp rate, or initial frequency, decreases with increasing temperature while the other two are fixed. The maximal temperature at which the reversal can happen increases with enlarging the system size. These phenomena are related to the facts that the energy barrier induced by anisotropy increases with the system volume, and the effective magnetization decreases with temperature. We also provide a set of optimal parameters for practical realization of our proposal. These findings may provide a way to realize low-cost and fast magnetization reversal with a wide operating temperature.
\end{abstract}


\maketitle

\section{Introduction}
Fast and energy-efficient magnetization reversal of magnetic nanoparticle draws much attention because of its application in non-volatile data storage device \cite{sun2000,woods2001,zitoun2002} and rapid information processing \cite{hillebrands2003}. For practical realization, a wide range of operating temperature is required which can be obtained by employing the high-anisotropy materials which belong to higher energy barrier \cite{mangin2006}. But the challenging issue is to find out the way to achieve fast magnetization reversal for high-anisotropy materials with minimal energy. In the early years, magnetization reversal was induced by a constant magnetic field \cite{hubert1998,sun2005} which requires the larger reversal time \cite{hubert1998} and it suffers from scalability and field localization issue for nano-device. The spin-polarized electric current directly or indirectly becomes a potential candidate to induce magnetization reversal through spin transfer torque (STT) and$\slash$or spin orbit torque (SOT) \cite{slonczewski1996,berger1996,tsoi1998,Katine2000,Waintal2000,sun2000a,suns2003,stiles2002,bazaliys2004,koch2004,wetzels2006,manchon2008,miron2010,miron2011,liu2012}.
However, in case of the STT-MRAM or SOT-MRAM based device fabrication, the requirement of a large current density is an obstacle since it generates Joule heat which limits the device durability and reliability \cite{grollier2003,morise2005,taniguchi2008,SUZUKI200993,zsun2006,wang2007,wange2008}.
Later on, people digress to utilize the microwave field of constant or time dependent frequency, either with or without a polarized electric current, to induce magnetization reversal \cite{bertotti2001,sunz2006,denisov2006,okamoto2008,zhu2010,thirion2003,rivkin2006,wangc2009,barros2011,barrose2013,tanaka2013,klughertzx2014}.

Recently, we proposed that a down-chirp microwave pulse (DCMWP) is capable of inducing subnanosecond magnetization reversal
with proper initial frequency $f_0$, chirp rate $\eta$ and field amplitude $H_\text{mw}$ at 0 temperature \cite{islam2018}.
However, finite temperature is ubiquitous in nature, and it is practically meaningful to study the effect of
finite temperature on DCMWP-induced magnetization reversal.
Indeed, the magnetization reversal is to overcome the energy barrier originated from the anisotropy, but the finite
temperature provides an isotropic background energy which makes the reversal easier.
There are several studies demonstrating the thermal effect assists magnetization reversal induced by magnetic field or electric current  \cite{koche2000,lia2004,de2017,iwata2018,liu2020}.
If the temperature is too high, there will
be a significant probability that the magnetization reverses simultaneously, which is undesired.
Therefore, it is meaningful to investigate how the optimal initial frequency, chirp rate and field amplitude
change with temperature, and the allowed temperature interval for DCMWP-induced magnetization reversal.
For device applications, the working temperature is usually the room temperature. So it is useful
to check whether the DCMWP-driven magnetization reversal is still valid at room temperature.
In this paper, we show that the thermal effect assists the DCMWP-induced magnetization reversal. Each of $f_0$, $\eta$ and $H_\text{mw}$ decreases with increasing temperature while the other two are fixed, until a maximal temperature above which the reversal is no longer valid. The maximal temperature increases with enlarging the system size and hence a wide operating temperature above room temperature is possible. Therefore, these findings may provide a way to realize low-cost and fast magnetization reversal with a wide operating temperature.

\section{Analytical model and method}

\begin{figure}
	\includegraphics[width=85mm]{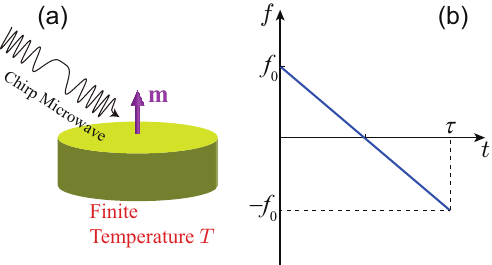}
\caption{\label{Fig1} (a) Schematic diagram of the system. The magnetization \(\mathbf{m}\)
is driven by a chirp microwave at finite temperature.
A down-chirp microwave field is applied onto a nanoparticle at finite temperature. (b) The frequency profile (sweeping from \(+f_0\) to  \(-f_0\)) of a down-chirp microwave. }
\end{figure}

We consider a single-domain ferromagnetic nanoparticle with uniaxial easy-axis anisotropy along $z$ axis at finite temperature $T$,  as shown in Fig. \ref{Fig1}(a). The magnetization direction is represented by a unit vector $\mathbf{m}$
with saturation magnetization $M_s$. When the nanoparticle is small, it can be regarded as a macrospin, as illustrated in the figure.
Nevertheless, we use smaller mesh size (from 0.5 nm$\times$0.5 nm$\times$0.5 nm to 2 nm$\times$2 nm$\times$2 nm,
depending on the edge length) to mimic the realistic condition.
The ground-state magnetization directions are all $\mathbf{m}$ parallel to $\hat{\mathbf{z}}$ and $-\hat{\mathbf{z}}$.

In the presence of a circularly polarized DCMWP and finite temperature, the magnetization dynamics is governed by the stochastic Landau Lifshitz Gilbert (sLLG) equation \cite{gilbert2004}
\begin{equation}
\frac{d\mathbf{m}}{dt}=-\gamma\mathbf{m}\times(\mathbf{H}_\text{eff}+\mathbf{h}_\text{th})+\alpha\mathbf{m}\times\frac{\partial\mathbf{m}}{\partial{t}}
\label{sllg}
\end{equation}
where  $\alpha$ and $\gamma$ are the Gilbert damping constant and gyromagnetic ratio respectively, and the effective field $(\mathbf{H}_\text{eff})$
comes from the microwave magnetic field $\mathbf{H}_\text{mw}$, the exchange field
$\frac{2A}{M_s}\nabla^2 \mathbf{m}$, and the easy-axis anisotropy field along $z$ direction $\mathbf{H}_k=H_km_z\hat{\mathbf{z}}$. $\mathbf{h}_\text{th}$ is the stochastic thermal field due to the finite temperature.
The thermal field follows the Gaussian process characterized by following relations \cite{Nowak2008}
\begin{equation}
 \begin{gathered}
 \langle h_{\text{th},ip}(t)\rangle = 0,\\
\langle h_{\text{th},ip}(t)h_{\text{th},jq}(t+\Delta t)\rangle = \frac{2\alpha k_{\text{B}}T }
{\gamma  M_s V}\delta_{ij}\delta_{pq}\delta (\Delta t),
 \end{gathered}
 \end{equation}
where $\langle\rangle$ denotes the ensemble average, $i$ and $j$ label the micromagnetic cells, $p$ and $q$ represent the Cartesian components of the thermal field, $V$ is the volume of one micromagnetic cell, and $k_{B}$ is the Boltzman constant. Since
we are considering a macrospin model, $V$ is the volume of the whole particle.

In the absence of external force and at zero temperature, the magnetization of nanoparticle has two stable states $\mathbf{m}\parallel \hat{\mathbf{z}}$ and $\mathbf{m}\parallel -\hat{\mathbf{z}}$ due to the easy axis. To reverse the magnetization from one equilibrium state to the other, it is previously reported \cite{islam2018} that a circularly polarized DCMWP which takes the form $\mathbf{H}_\text{mw}=H_\text{mw}\left[\cos\phi(t)\hat{\mathbf{x}}+\sin\phi(t)\hat{\mathbf{y}}\right]\label{app}$ can induce fast reversal at zero temperature limit.  $\phi(t)$ is the phase giving the instantaneous frequency of DCMWP $f(t)\equiv \frac{1}{2\pi} \frac{\mathrm{d}\phi}{\mathrm{d}t}$, the frequency linearly decreases from $f_0$ to $-f_0$ with a chirp rate $\eta$ (with unit of ns$^{-2}$)
as $f=f_0-\eta t$, as shown in Fig. \ref{Fig1}(b). The phase is thus $\phi(t)=2\pi (f_{0}{t}-\frac{\eta}{2}t^2)$, and the duration of the microwave pulse is $\tau=\frac{2f_{0}}{\eta}$.
Subnanosecond magnetization reversal is achieved with the physical picture that the DCMWP triggers stimulated microwave absorptions (emissions) by (from) the spin before (after) it crosses over the energy barrier at $T = 0$ \cite{islam2018}.

In this study, the following parameters are selected from the representative experiments on microwave-driven magnetization reversal as $M_{s}=10^{6}$ A/m, ${H_\text{k}}$ = $0.75$ T, $\gamma$ = $1.76\times 10^{11}$ rad/T/s, $A$ = $13 \times 10^{−12}$ J/m, and $\alpha= 0.01$.
We use MuMax3 Package \cite{vansteenkiste2014} to numerically solve stochastic LLG equation choosing adaptive Heun
solver. To rule out the effect of shape anisotropy, we consider cubic nanoparticles with edge length $a$ throughout this paper.
$a=2,4,8,12$ nm are studied. Considering the time-efficiency as well as stability \cite{islam2019}, we use the time
step of $10^{−15}$ s for small volume ($2\times 2 \times 2 \text{ nm}^3$), and $10^{−14}$ s for other larger volumes.
Each numerical result is obtained from the average of 12 random tests with different random seeds. We define the switching time as
the time at which the magnetization reaches $m_z=-0.7$, considering practical requirement.

\section{Numerical Results}

\begin{figure*}
\begin{center}
    \includegraphics[width=0.95\textwidth]{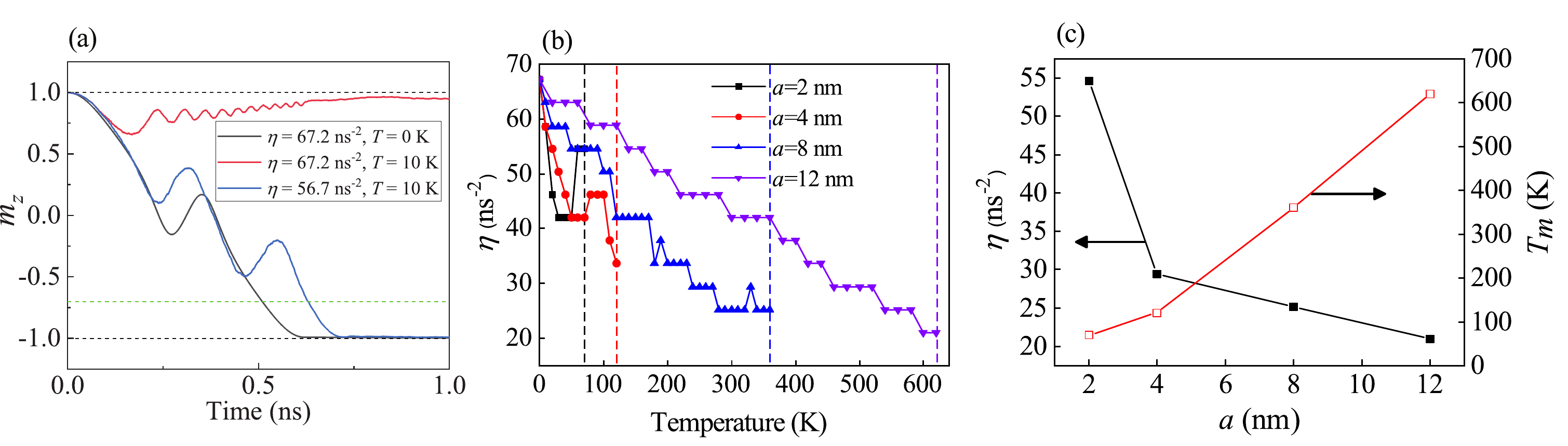}
\end{center}
	
\caption{\label{Fig2} (a) The time evolution of $m_z$ of $a=4$ nm driven by the DCMWP with ${f_0}$ = $21$ GHz, $H_\text{mw} = 0.045$ T and  optimal chirp rates $\eta$ for different $T$. (b) Optimal chirp rate $\eta$ as a function of $T$ for different cubic nanoparticles edge length $a$. (c) Estimated optimal chirp rate $\eta$ (solid squares) and maximal temperature $T_m$ (open squares) are the function of $a$. }
\end{figure*}

We first focus on a $4\times4\times 4$ nm$^3$ nanoparticle and study the effect of temperature on the DCMWP-driven magnetization reversal. We keep the initial frequency $f_0=21$ GHz, and the microwave field $H_\text{mw}=0.045$ T (which are same as the zero-temperature parameters obtained in \cite{islam2018}) fixed, and try to find the optimal chirp rate $\eta$ (i.e. the chirp rate that leads to fastest reversal). The zero-temperature optimal chirp rate $\eta_0$ is 67.2 ns$^{-2}$ according to \cite{islam2018}).

The temporal evolution of $m_z$ at this optimal chirp rate and at 0 K is shown by the black line in Fig. \ref{Fig2}(a).
However, when the temperature becomes finite ($T=10$ K), the same chirp rate can no longer drive the magnetization reversal, as shown
by the red line. By tuning the chirp rate $\eta$, we find that $\eta=56.7$ ns$^{-2}$ leads to the fastest reversal for $T=10$ K,
as shown by the blue line. Thus, we find that the optimal chirp rate is temperature-dependent, which is an issue has to be considered
in device applications.
For the 4-nm cube, when $T>120$ K, the reversal to $m_z=-0.7$ is no longer valid. We define this highest temperature that the
DCMWP-driven reversal is possible as $T_m$. $T_m=120$ K for the 4-nm cube is not practically useful. To increase $T_m$ so that
it's higher than the room temperature, we try to enlarge the volume of the sample. Because the anisotropy energy that stabilize $\mathbf{m}$ towards the north and south poles is $E_a=KV$ which is proportional to the volume but the thermal energy $E_t=k_BT$ is independent of
the volume \cite{koche2000,lia2004,de2017,iwata2018,liu2020}, increasing the sample volume helps enhancing the stability so that $T_m$ may be higher. We study $a=2$, 4, 8, 12 nm cubic samples, simulate the DCMWP-driven dynamics at different chirp rates, and find the optimal chirp rates for different temperatures. Figure \ref{Fig2}(b) shows the optimal chirp rates $\eta$ (in a step size of 4.2 ns$^{-2}$) as the function of temperature for different sample sizes. $\eta$ shows an overall decreasing trend with increasing temperature, despite some fluctuations which may due to the complex dynamics induced by the stochastic field and chirp field. The vertical dashed line is the maximal temperature $T_m$ for each size, which increases with the sample size. For the 12-nm sample, $T_m$ is above room temperature which means DCMWP-driven magnetization reversal is possible at room temperature and useful in device applications. To be more explicit, we plot $T_m$ and the optimal chirp rate at $T_m$ [$\eta(T_m)$] versus the sample volume $V=a^3$.
As expected, the $T_m$ increases with $V$, while $\eta(T_m)$ decreases with $V$. Since the reversal time is closed to the pulse duration
$\tau=\frac{2f_0}{\eta}$, the reversal time increases with the temperature, as shown in the Fig. \ref{Fig2}(c).

\begin{figure}[pos=ht]
	\includegraphics[width=75mm]{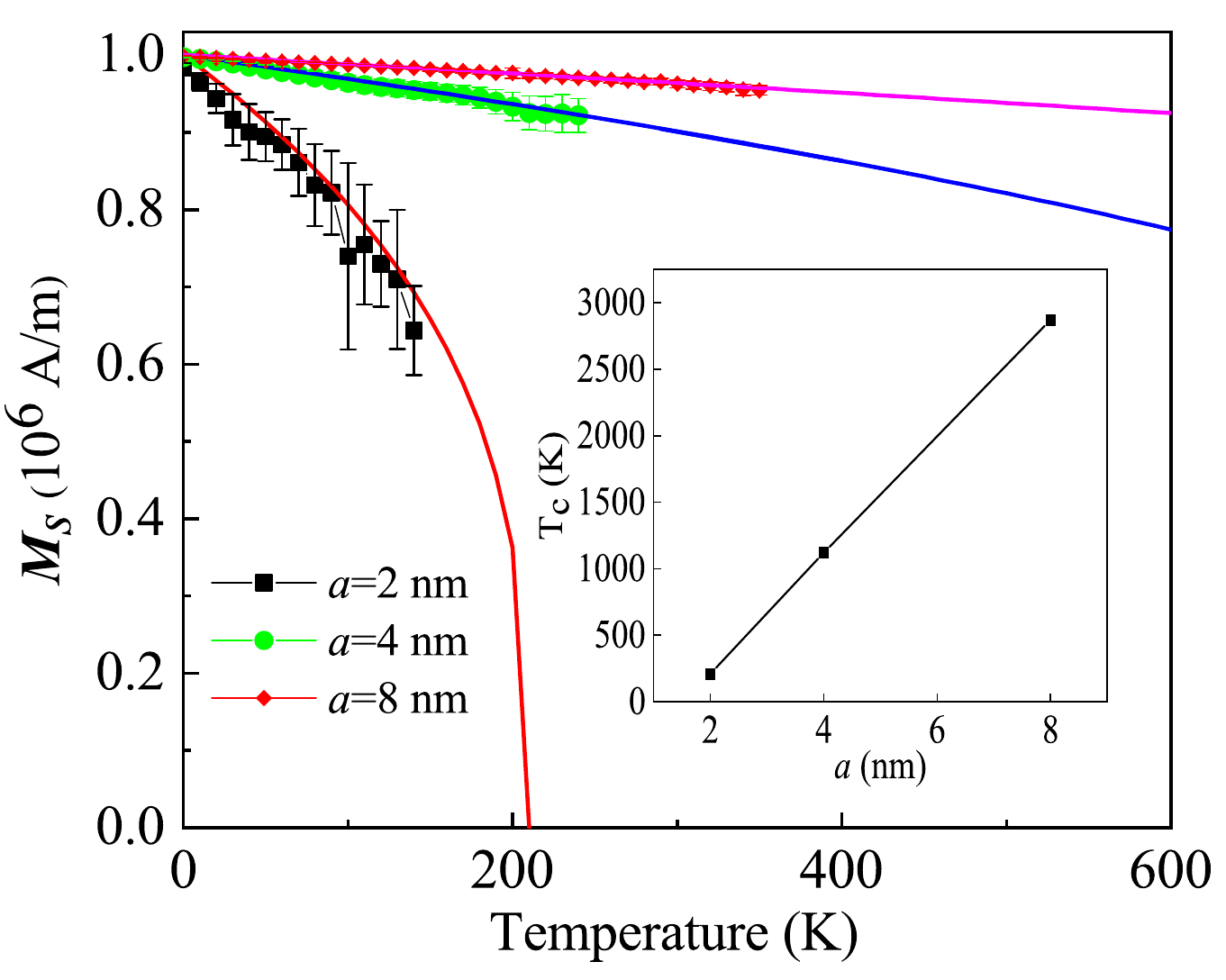}
\caption{\label{Fig3} $M_{s}$ is a function of $T$ for different cubic nanoparticles edge, $a$. Inset shows the Curie temperature, $T_c$ as a function of the edge length, $a$. }
\end{figure}

To understand the reason why the optimal chirp rate decreases with temperature, we recall the basic knowledge that the effective saturation magnetization $M_s$ decreases with temperature due to the spin-wave excitation \cite{Kittelbook}. We numerically calculate the temperature dependence of saturation magnetization $M_s$ by taking a long-time average of $M_z$ without any external driving forces. It is found that  $M_s$ decreases with $T$ rapidly for smaller volume but decreases slowly for larger volume. The $M_s-T$ relation can be well fitted by the well-known power law ${M_s}(T)={M_{s}(0)}\times(1-\frac{T}{T_c})^{1/3}$ where $T_c$ is the Curie temperature.
By extrapolating the $M_s-T$ curves to $M_s=0$,  we estimated the Curie temperature, as shown in the insets of the Fig. \ref{Fig3}. The Curie temperature increases with the volume as expected. The reduced effective $M_s$ leads to a reduced intrinsic characteristic frequency
$\gamma M_s$. Consequently, all the dynamics slow down (as if the time scale of the dynamics is ``expanded") and the reversal time as well as the inverse of chirp rate increases correspondingly.

\begin{figure}[pos=ht]
    \centering
	\includegraphics[width=75mm]{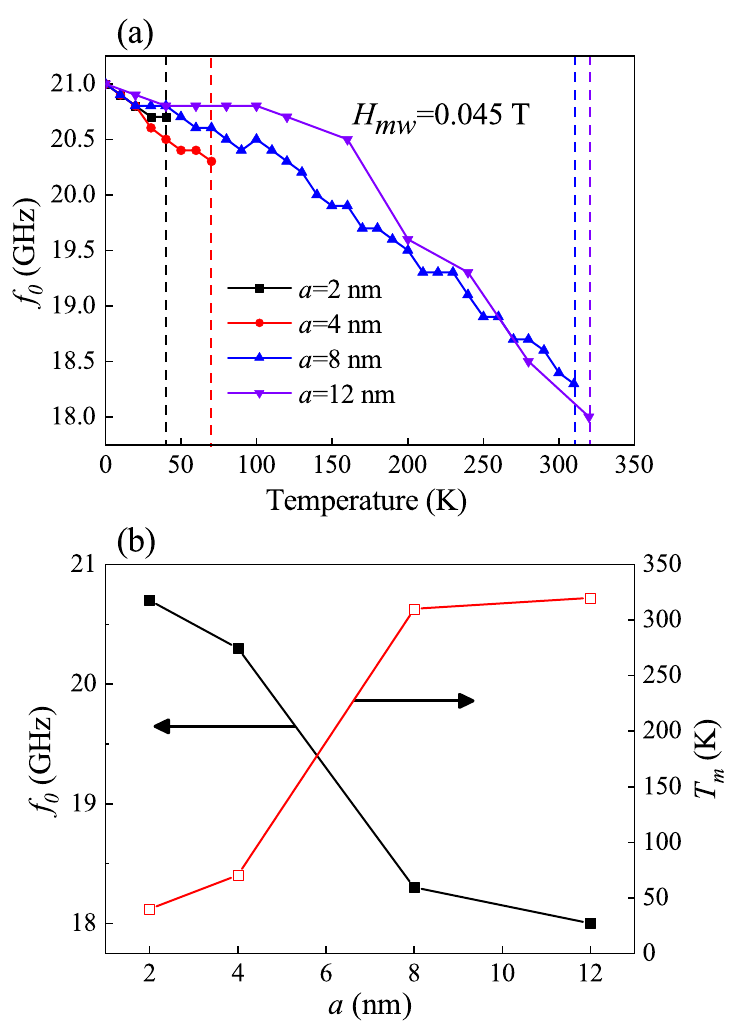}
\caption{\label{Fig4} (a) Temperature $T$ dependence of $f_0$ while $H_\text{mw}=0.045$ T and the optimal $\eta$ at corresponding $T$ are fixed. (b) The minimal $f_0$ (solid squares) at $T_m$ and the maximal temperature $T_m$ (open squares) as the function of the edge length $a$.}
\end{figure}

Due to the reduced effective $M_s$ and ``expanded" time, we can expect that the optimal initial frequency $f_0$  should also decrease with temperature.  Purposely, by keeping the microwave amplitude $H_\text{mw}=0.045$ T and the optimal chirp rate $\eta$ at corresponding temperature fixed, we examine the temperature effect on $f_0$ of DCMWP for the sample $a=4$ nm cube.  It is observed that the minimal $f_0$ deceases with temperature $T$ is shown by red line in the Fig. \ref{Fig4} (a). For the 4-nm cubic sample, the minimal $f_0$ is 20.3 GHz at the maximal temperature  which is still far below the room temperature. To reduce $f_0$ and increase $T_m$ further, we study the magnetization reversal with enlarging the sample volume. Fig. \ref{Fig4} (a) shows that the minimal $f_0$ decreases with $T$ as well as sample volume which is consistent with the reduction of effective $M_s$. For smaller sample, $f_0$ decreases with $T$ rapidly but for larger volume, decreases slowly. The vertical dashed lines indicate the minimal $f_0$ at $T_m$. To be more explicit, the minimal $f_0$ at $T_m$ i.e., [$f_0(T_m)$] and $T_m$ are plotted as a function of the sample volume. The minimal $f_0$ at $T_m$ decreases (black line) but $T_m$ increases (red line) with the sample volume shown in Fig. \ref{Fig4}(b). Specifically, for 12-nm sample at above the room temperature $T_m= 320$ K, with  $\eta=42$ ns$^{-2}$ and $H_\text{mw}=0.045$ T, $f_0$ significantly reduces to 18 GHz.

\begin{figure}[pos=ht]
   \centering
	\includegraphics[width=75mm]{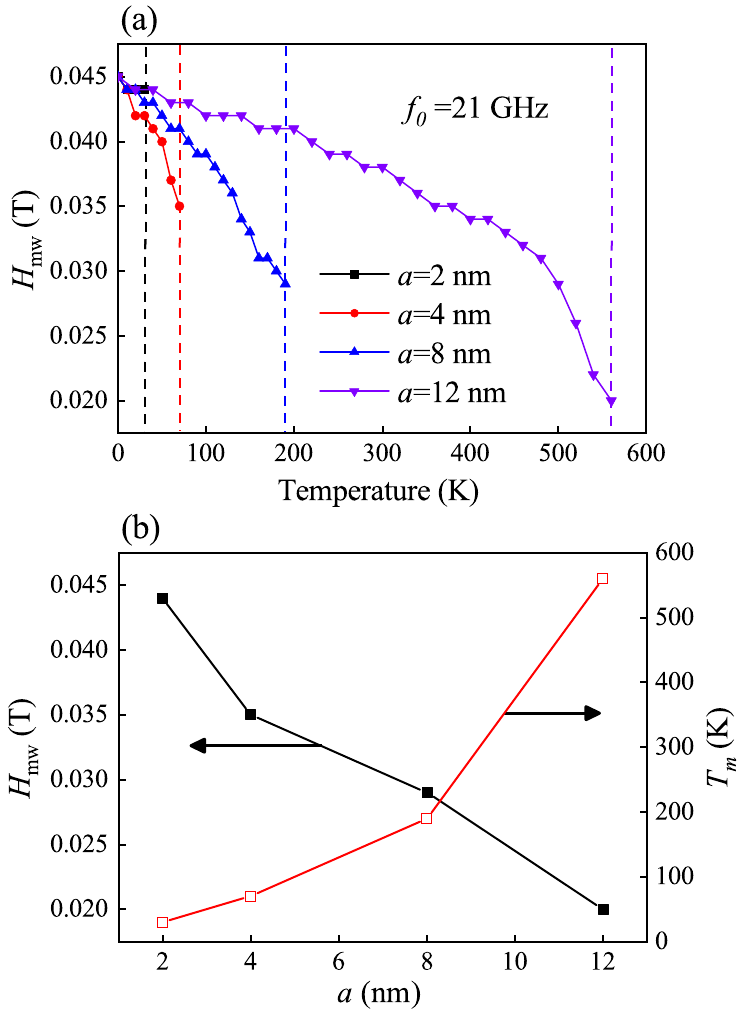}
\caption{\label{Fig5}  (a) Temperature $T$ dependence of $H_\text{mw}$ while $f_0 = 21$ GHz and the optimal $\eta$ at corresponding $T$ are fixed.   (b) The minimal $H_\text{mw}$ (solid squares) at $T_m$  and the maximal temperature $T_m$ (open squares) as a function of the cubic edge $a$.}
\end{figure}
Later on, by keeping the $f_0 = 21$ GHz and the optimal chirp rate $\eta$ at corresponding $T$ fixed, we investigate the temperature dependence of the minimally $H_\text{mw}$ of DCMWP required for different sample volumes (2 nm, 4 nm, 8 nm and 12 nm). In this case we also found that the minimal $H_\text{mw}$ decreases with the temperature $T$, as shown in Fig. \ref{Fig5}(a). This is because the finite temperature  provides an isotropic background energy which assists the magnetization reversal. Similarly, $H_\text{mw}$ decreases with $T$ up to a maximal temperature $T_m$, indicated by vertical dashed lines.  Figure \ref{Fig5}(b) demonstrates explicitly how the maximal working temperature $T_m$ and the minimal $H_\text{mw}$ at $T_m$ [$H_\text{mw}(T_m)$] depend on the sample size. decreases (black line) $T_m$ increases (red symbols, right axis) and $H_\text{mw}(T_m)$ (black symbols, left axis) decreases with the sample size. 

Therefore, for 12-nm (or larger) cubic sample, the working temperature can be higher than room temperature, which is practically meaningful. So we search the full parameter space of the DCMWP to find optimal parameters at room temperature: $H_\text{mw}\sim0.038$ T,  $f_0 \sim 19$ GHz and  $\eta\sim$ 38 ns $^{-2}$, which gives a fast magnetization reversal time 2 ns. For device applications, one can simulate the reversal dynamics for an actual nano-particle and find the optimal parameters, then this set of parameters can be fixed and used for all the same nano-particles.

\section{Discussion and Conclusions}
With zero temperature  \cite{islam2018}, DCMWP-driven subnanosecond magnetization reversal has been demenstrated in zero-temperature limit \cite{islam2018}. In this study, finite temperature effects are included, and  the DCMWP-driven fast magnetization reversal is still valid. Since the temperature effectively reduces the energy barrier and expands the time scale, the optimal chirp rate, the optimal initial frequency and the minimal field amplitude decrease with temperature. For the same reason, when the temperature is too high, the magnetization loses its stability, so there is a maximal working temperature for the DCMWP-driven regime. The maximal working temperature can be raised by enlarging the sample volume. In our simulations, 12-nm cubic sample has a maximal working temperature of 320 K, which is slightly higher than the room temperature, indicating that the DCMWP-driven magnetization reversal is practically possible at room temperature for nano-particles larger than $12^3$ nm$^3$. For the 12-nm sample, we find a set of optimal parameters at room temperature, i.e., $H_\text{mw}\sim0.038$ T,  $f_0 \sim 19$ GHz and  $\eta\sim$ 38 ns $^{-2}$ which are useful in device application. Since the macrospin model is valid up to 30-nm-diameter nano-particles \cite{iwata2018}, it is expected that our proposal works for larger nano-particles practically. There are the recent technologies  \cite{cai2013,cai2010} to generate the required microwave chirp pulse.  Therefore, these findings may provide a way to realize low-cost and fast magnetization reversal with a wide operating temperature.

\section{Acknowledgements}
This work was supported by the Khulna University Research Cell (Grant No. KU/RC-04/2000-232).
X. S. W. acknowledges the support from the Natural Science Foundation of China (NSFC) (Grant No. 11804045).

\bibliographystyle{elsarticle-num}

\bibliography{main}

\begin{thebibliography}{10}
\expandafter\ifx\csname url\endcsname\relax
  \def\url#1{\texttt{#1}}\fi
\expandafter\ifx\csname urlprefix\endcsname\relax\def\urlprefix{URL }\fi
\expandafter\ifx\csname href\endcsname\relax
  \def\href#1#2{#2} \def\path#1{#1}\fi

\bibitem{sun2000}
S.~Sun, C.~B. Murray, D.~Weller, L.~Folks, A.~Moser, Monodisperse fept
  nanoparticles and ferromagnetic fept nanocrystal superlattices, science
  287~(5460) (2000) 1989--1992.

\bibitem{woods2001}
S.~Woods, J.~Kirtley, S.~Sun, R.~Koch, Direct investigation of
  superparamagnetism in co nanoparticle films, Physical review letters 87~(13)
  (2001) 137205.

\bibitem{zitoun2002}
D.~Zitoun, M.~Respaud, M.-C. Fromen, M.~J. Casanove, P.~Lecante, C.~Amiens,
  B.~Chaudret, Magnetic enhancement in nanoscale corh particles, Physical
  review letters 89~(3) (2002) 037203.

\bibitem{hillebrands2003}
B.~Hillebrands, K.~Ounadjela, Spin dynamics in confined magnetic structures I
  \& II, Vol.~83, Springer Science \& Business Media, 2003.

\bibitem{mangin2006}
S.~Mangin, D.~Ravelosona, J.~Katine, M.~Carey, B.~Terris, E.~E. Fullerton,
  Current-induced magnetization reversal in nanopillars with perpendicular
  anisotropy, Nature materials 5~(3) (2006) 210--215.

\bibitem{hubert1998}
A.~Hubert, R. scha fer, magnetic domains: the analysis of magnetic
  microstructures (1998).

\bibitem{sun2005}
Z.~Sun, X.~Wang, Fast magnetization switching of stoner particles: A nonlinear
  dynamics picture, Physical Review B 71~(17) (2005) 174430.

\bibitem{slonczewski1996}
J.~C. Slonczewski, et~al., Current-driven excitation of magnetic multilayers,
  Journal of Magnetism and Magnetic Materials 159~(1) (1996) L1.

\bibitem{berger1996}
L.~Berger, Emission of spin waves by a magnetic multilayer traversed by a
  current, Physical Review B 54~(13) (1996) 9353.

\bibitem{tsoi1998}
M.~Tsoi, A.~Jansen, J.~Bass, W.-C. Chiang, M.~Seck, V.~Tsoi, P.~Wyder,
  Excitation of a magnetic multilayer by an electric current, Physical Review
  Letters 80~(19) (1998) 4281.

\bibitem{Katine2000}
J.~A. Katine, F.~J. Albert, R.~A. Buhrman, E.~B. Myers, D.~C. Ralph,
  Current-driven magnetization reversal and spin-wave excitations in co $/$cu
  $/$co pillars, Phys. Rev. Lett. 84 (2000) 3149--3152.

\bibitem{Waintal2000}
X.~Waintal, E.~B. Myers, P.~W. Brouwer, D.~C. Ralph, Role of spin-dependent
  interface scattering in generating current-induced torques in magnetic
  multilayers, Phys. Rev. B 62 (2000) 12317--12327.

\bibitem{sun2000a}
J.~Z. Sun, Spin-current interaction with a monodomain magnetic body: A model
  study, Physical Review B 62~(1) (2000) 570.

\bibitem{suns2003}
J.~Sun, Spintronics gets a magnetic flute, Nature 425~(6956) (2003) 359--360.

\bibitem{stiles2002}
M.~D. Stiles, A.~Zangwill, Anatomy of spin-transfer torque, Physical Review B
  66~(1) (2002) 014407.

\bibitem{bazaliys2004}
Y.~B. Bazaliy, B.~Jones, S.-C. Zhang, Current-induced magnetization switching
  in small domains of different anisotropies, Physical Review B 69~(9) (2004)
  094421.

\bibitem{koch2004}
R.~Koch, J.~Katine, J.~Sun, Time-resolved reversal of spin-transfer switching
  in a nanomagnet, Physical review letters 92~(8) (2004) 088302.

\bibitem{wetzels2006}
W.~Wetzels, G.~E. Bauer, O.~N. Jouravlev, Efficient magnetization reversal with
  noisy currents, Physical review letters 96~(12) (2006) 127203.

\bibitem{manchon2008}
A.~Manchon, S.~Zhang, Theory of nonequilibrium intrinsic spin torque in a
  single nanomagnet, Physical Review B 78~(21) (2008) 212405.

\bibitem{miron2010}
I.~M. Miron, G.~Gaudin, S.~Auffret, B.~Rodmacq, A.~Schuhl, S.~Pizzini,
  J.~Vogel, P.~Gambardella, Current-driven spin torque induced by the rashba
  effect in a ferromagnetic metal layer, Nature materials 9~(3) (2010)
  230--234.

\bibitem{miron2011}
I.~M. Miron, K.~Garello, G.~Gaudin, P.-J. Zermatten, M.~V. Costache,
  S.~Auffret, S.~Bandiera, B.~Rodmacq, A.~Schuhl, P.~Gambardella, Perpendicular
  switching of a single ferromagnetic layer induced by in-plane current
  injection, Nature 476~(7359) (2011) 189--193.

\bibitem{liu2012}
L.~Liu, C.-F. Pai, Y.~Li, H.~Tseng, D.~Ralph, R.~Buhrman, Spin-torque switching
  with the giant spin hall effect of tantalum, Science 336~(6081) (2012)
  555--558.

\bibitem{grollier2003}
J.~Grollier, V.~Cros, H.~Jaffres, A.~Hamzic, J.-M. George, G.~Faini, J.~B.
  Youssef, H.~Le~Gall, A.~Fert, Field dependence of magnetization reversal by
  spin transfer, Physical Review B 67~(17) (2003) 174402.

\bibitem{morise2005}
H.~Morise, S.~Nakamura, Stable magnetization states under a spin-polarized
  current and a magnetic field, Physical Review B 71~(1) (2005) 014439.

\bibitem{taniguchi2008}
T.~Taniguchi, H.~Imamura, Critical current of spin-transfer-torque-driven
  magnetization dynamics in magnetic multilayers, Physical Review B 78~(22)
  (2008) 224421.

\bibitem{SUZUKI200993}
Y.~Suzuki, A.~A. Tulapurkar, C.~Chappert, Spin-injection phenomena and
  applications, in: Nanomagnetism and Spintronics, Elsevier, 2009, pp. 93--153.

\bibitem{zsun2006}
Z.~Sun, X.~Wang, Theoretical limit of the minimal magnetization switching field
  and the optimal field pulse for stoner particles, Physical review letters
  97~(7) (2006) 077205.

\bibitem{wang2007}
X.~Wang, Z.~Sun, Theoretical limit in the magnetization reversal of stoner
  particles, Physical review letters 98~(7) (2007) 077201.

\bibitem{wange2008}
X.~Wang, P.~Yan, J.~Lu, C.~He, Euler equation of the optimal trajectory for the
  fastest magnetization reversal of nano-magnetic structures, EPL (Europhysics
  Letters) 84~(2) (2008) 27008.

\bibitem{bertotti2001}
G.~Bertotti, C.~Serpico, I.~D. Mayergoyz, Nonlinear magnetization dynamics
  under circularly polarized field, Physical Review Letters 86~(4) (2001) 724.

\bibitem{sunz2006}
Z.~Sun, X.~Wang, Strategy to reduce minimal magnetization switching field for
  stoner particles, Physical Review B 73~(9) (2006) 092416.

\bibitem{denisov2006}
S.~I. Denisov, T.~V. Lyutyy, P.~H{\"a}nggi, K.~N. Trohidou, Dynamical and
  thermal effects in nanoparticle systems driven by a rotating magnetic field,
  Physical Review B 74~(10) (2006) 104406.

\bibitem{okamoto2008}
S.~Okamoto, N.~Kikuchi, O.~Kitakami, Magnetization switching behavior with
  microwave assistance, Applied Physics Letters 93~(10) (2008) 102506.

\bibitem{zhu2010}
J.-G. Zhu, Y.~Wang, Microwave assisted magnetic recording utilizing
  perpendicular spin torque oscillator with switchable perpendicular
  electrodes, IEEE Transactions on Magnetics 46~(3) (2010) 751--757.

\bibitem{thirion2003}
C.~Thirion, W.~Wernsdorfer, D.~Mailly, Switching of magnetization by nonlinear
  resonance studied in single nanoparticles, Nature materials 2~(8) (2003)
  524--527.

\bibitem{rivkin2006}
K.~Rivkin, J.~B. Ketterson, Magnetization reversal in the anisotropy-dominated
  regime using time-dependent magnetic fields, Applied physics letters 89~(25)
  (2006) 252507.

\bibitem{wangc2009}
Z.~Wang, M.~Wu, Chirped-microwave assisted magnetization reversal, Journal of
  Applied Physics 105~(9) (2009) 093903.

\bibitem{barros2011}
N.~Barros, M.~Rassam, H.~Jirari, H.~Kachkachi, Optimal switching of a
  nanomagnet assisted by microwaves, Physical Review B 83~(14) (2011) 144418.

\bibitem{barrose2013}
N.~Barros, H.~Rassam, H.~Kachkachi, Microwave-assisted switching of a
  nanomagnet: Analytical determination of the optimal microwave field, Physical
  Review B 88~(1) (2013) 014421.

\bibitem{tanaka2013}
T.~Tanaka, Y.~Otsuka, Y.~Furomoto, K.~Matsuyama, Y.~Nozaki, Selective
  magnetization switching with microwave assistance for three-dimensional
  magnetic recording, Journal of Applied Physics 113~(14) (2013) 143908.

\bibitem{klughertzx2014}
G.~Klughertz, P.-A. Hervieux, G.~Manfredi, Autoresonant control of the
  magnetization switching in single-domain nanoparticles, Journal of Physics D:
  Applied Physics 47~(34) (2014) 345004.

\bibitem{islam2018}
M.~T. Islam, X.~Wang, Y.~Zhang, X.~Wang, Subnanosecond magnetization reversal
  of a magnetic nanoparticle driven by a chirp microwave field pulse, Physical
  Review B 97~(22) (2018) 224412.

\bibitem{koche2000}
R.~H. Koch, G.~Grinstein, G.~Keefe, Y.~Lu, P.~Trouilloud, W.~Gallagher,
  S.~Parkin, Thermally assisted magnetization reversal in submicron-sized
  magnetic thin films, Physical review letters 84~(23) (2000) 5419.

\bibitem{lia2004}
Z.~Li, S.~Zhang, Thermally assisted magnetization reversal in the presence of a
  spin-transfer torque, Physical Review B 69~(13) (2004) 134416.

\bibitem{de2017}
J.~De~Vries, T.~Bolhuis, L.~Abelmann, Temperature dependence of the energy
  barrier and switching field of sub-micron magnetic islands with perpendicular
  anisotropy, New journal of physics 19~(9) (2017) 093019.

\bibitem{iwata2018}
J.~M. Iwata-Harms, G.~Jan, H.~Liu, S.~Serrano-Guisan, J.~Zhu, L.~Thomas, R.-Y.
  Tong, V.~Sundar, P.-K. Wang, High-temperature thermal stability driven by
  magnetization dilution in cofeb free layers for spin-transfer-torque magnetic
  random access memory, Scientific reports 8~(1) (2018) 1--7.

\bibitem{liu2020}
W.~Liu, B.~Cheng, S.~Ren, W.~Huang, J.~Xie, G.~Zhou, H.~Qin, J.~Hu, Thermally
  assisted magnetization control and switching of dy3fe5o12 and tb3fe5o12
  ferrimagnetic garnet by low density current, Journal of Magnetism and
  Magnetic Materials (2020) 166804.

\bibitem{gilbert2004}
T.~L. Gilbert, A phenomenological theory of damping in ferromagnetic materials,
  IEEE transactions on magnetics 40~(6) (2004) 3443--3449.

\bibitem{Nowak2008}
D.~Hinzke, N.~Kazantseva, U.~Nowak, O.~N. Mryasov, P.~Asselin, R.~W. Chantrell,
  \href{https://link.aps.org/doi/10.1103/PhysRevB.77.094407}{Domain wall
  properties of fept: From bloch to linear walls}, Phys. Rev. B 77 (2008)
  094407.
\newblock \href {https://doi.org/10.1103/PhysRevB.77.094407}
  {\path{doi:10.1103/PhysRevB.77.094407}}.
\newline\urlprefix\url{https://link.aps.org/doi/10.1103/PhysRevB.77.094407}

\bibitem{vansteenkiste2014}
A.~Vansteenkiste, J.~Leliaert, M.~Dvornik, M.~Helsen, F.~Garcia-Sanchez,
  B.~Van~Waeyenberge, The design and verification of mumax3, AIP advances
  4~(10) (2014) 107133.

\bibitem{islam2019}
M.~T. Islam, X.~Wang, X.~Wang, Thermal gradient driven domain wall dynamics,
  Journal of Physics: Condensed Matter 31~(45) (2019) 455701.

\bibitem{Kittelbook}
C.~Kittel,
  \href{http://www.amazon.com/Introduction-Solid-Physics-Charles-Kittel/dp/047141526X/ref=dp_ob_title_bk}{Introduction
  to Solid State Physics}, 8th Edition, Wiley, 2004.
\newline\urlprefix\url{http://www.amazon.com/Introduction-Solid-Physics-Charles-Kittel/dp/047141526X/ref=dp_ob_title_bk}

\bibitem{cai2013}
L.~Cai, D.~A. Garanin, E.~M. Chudnovsky, Reversal of magnetization of a
  single-domain magnetic particle by the ac field of time-dependent frequency,
  Physical Review B 87~(2) (2013) 024418.

\bibitem{cai2010}
L.~Cai, E.~M. Chudnovsky, Interaction of a nanomagnet with a weak
  superconducting link, Physical Review B 82~(10) (2010) 104429.

\end{thebibliography}

\end{document}